\documentclass[epj, twocolumn]{webofc}
\usepackage[varg]{txfonts}   

\newcommand{\dEdx}{\ensuremath{\mathrm{d}E/\mathrm{d}x}}
\newcommand{\PbPb}{\ensuremath{\mbox{Pb--Pb}}}
\newcommand{\pp}{\ensuremath{\mathrm {p\kern-0.05em p}}}
\usepackage{siunitx}
\DeclareSIUnit\clight{$c$}

\wocname{MPGD2015}
\woctitle{4th Conference on Micro-Pattern Gaseous Detectors}

\begin{document}

\title{Study of the d\textit{E}/d\textit{x} resolution of a GEM Readout Chamber prototype for the upgrade of the ALICE TPC}
\author{Andreas Mathis\inst{1,2}\fnsep\thanks{\email{andreas.mathis@ph.tum.de}} on behalf of the ALICE collaboration}
\institute{Physik Department E62, Technische Universit{\"a}t M{\"u}nchen, 85748 Garching, Germany \and Excellence Cluster ’Origin and Structure of the Universe’, 85748 Garching, Germany}
\abstract{The ALICE Collaboration is planning a major upgrade of its central barrel detectors to be able to cope with the increased LHC luminosity beyond 2020. For the TPC, this implies a replacement of the currently used gated MWPCs (Multi-Wire Proportional Chamber) by GEM (Gas Electron Multiplier) based readout chambers.

In order to prove, that the present particle identification capabilities via measurement of the specific energy loss are retained after the upgrade, a prototype of the ALICE IROC (Inner Readout Chamber) has been evaluated in a test beam campaign at the CERN PS. The \dEdx{} resolution of the prototype has been proven to be fully compatible with the current MWPCs.}

\maketitle

\section{Introduction}
\label{intro}
ALICE (A Large Ion Collider Experiment) \cite{ALICE} is a heavy-ion experiment designed to study the physics of the high-density, high-temperature phase of strongly interacting matter, and in particular the properties of the Quark–Gluon Plasma (QGP) in proton–proton, proton–nucleus and nucleus–nucleus collisions at the CERN Large Hadron Collider (LHC). 

The second generation of LHC heavy-ion studies beyond 2020 (in LHC RUN3 after the so called Long Shutdown 2) will be enabled by a significant increase of the LHC instant luminosity up to $6\times10^{27}$ \si{cm^{-2}s^{-1}}, thus allowing for the study of rare probes and their coupling to the medium \cite{ALICE_UPGR}. 

In order to fully exploit this increased luminosity, a continuous readout of the ALICE detector at an interaction rate of \SI{50}{\kilo\hertz} is foreseen.

\section{ALICE TPC - status and perspectives}
\label{sec-ALICE_TPC}
The ALICE Time Projection Chamber \cite{NIMA_TPC} is the main device for charged-particle tracking, momentum measurement and particle identification (PID) in the central barrel of the ALICE experiment. 
It covers a transverse momentum range $0.1 < p_{\mathrm{T}} < 100$\,GeV/$c$ within the pseudo rapidity range $\lvert\eta\rvert \leq 0.9$ and full azimuth.

The TPC consists of a hollow cylindrical barrel with the cathode in its center and two readout planes on its sides. The drift field of \SI{400}{\volt/\centi\meter} is precisely defined by the cathode and a cylindrical field cage degrading the potential from \SI{100}{\kilo\volt} at the cathode to nearly ground potential close to the readout chambers. 
The active volume has an inner radius of \SI{84.8}{\centi\meter}, an outer radius of \SI{246.6}{\centi\meter} and an overall length along the beam axis of \SI{499.4}{\centi\meter}. With an overall volume of about \SI{90}{\cubic\meter}, it is the largest detector of its kind in the world. 

\begin{figure}
\centering
\includegraphics[width=0.5\textwidth]{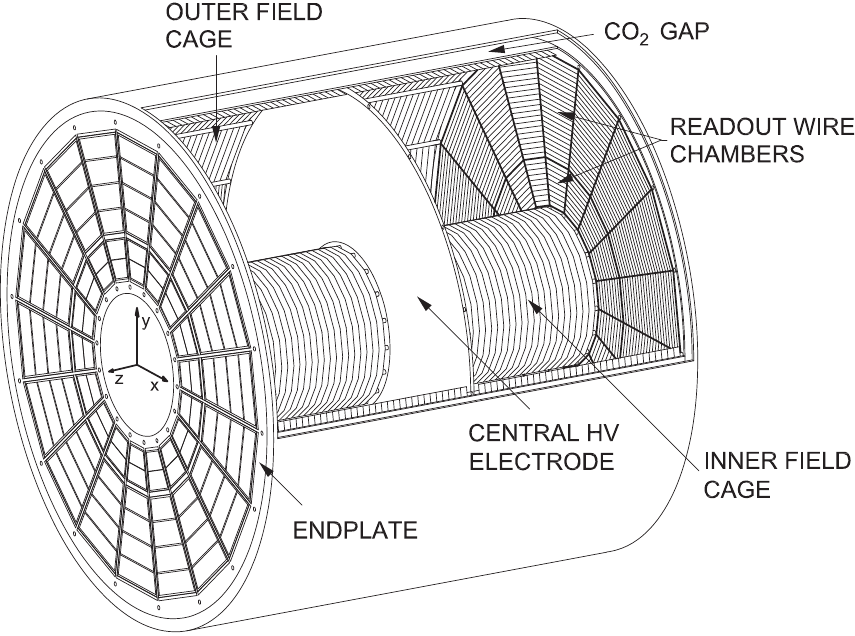}
\caption{Schematic drawing of the ALICE TPC \cite{NIMA_TPC}.}
\label{img:TPC}
\end{figure}

A schematic layout of the TPC is shown in Fig. \ref{img:TPC} with its most important components depicted.
The two endplates are azimuthally segmented into 18 sectors, each covering \SI{20}{\degree}. A further division of each sector into Inner (IROC) and Outer Readout Chamber (OROC) is motivated by the different requirements for the readout chambers as a function of the radius due to the radial dependence of the track density. The design of the readout chambers is based on a gated MWPC with cathode pad readout operated at a gain of 7000-8000.

The necessity to operate the TPC in a gated mode arises from the imperative to minimize distortions of the drift field by ions released from the amplification region. However, this sets severe constraints on the maximally achievable readout rate, as it allows for no more than \SI{3.5}{\kilo\hertz} in \pp{}. In \PbPb{}, the readout rate is further limited to $\sim$\SI{300}{\hertz} by the currently used readout electronics. 

In RUN3 however, the interaction rate will be increased to \SI{50}{\kilo\hertz}, implying an average event pileup of 5 \cite{ALICE_UPGR}. A triggered operation of the TPC with a gating grid will then no longer be possible, as it would cause unacceptable losses of data. Hence, a continuous operation of the TPC is mandatory, as is the minimization of the ion backflow with other means than a gating grid.

The baseline solution for the TPC upgrade consists of a stack of four large-size GEM \cite{GEM} foils as amplification stage, containing both Standard (S, \SI{140}{\micro\meter}) and Large Pitch (LP, \SI{280}{\micro\meter}) GEM foils arranged in the order S-LP-LP-S. The high voltage configuration applied to the stack has been optimized in order to fulfill the design specifications in terms of ion backflow, energy resolution and stable operation under LHC conditions \cite{TDR_TPCU-add}.

\section{d\textit{E}/d\textit{x} resolution studies with an Inner Readout Chamber}
\label{sec-IROC}
In order to validate the performance of a large-size detector equipped with a stack of four GEMs in terms of \dEdx{} resolution, a prototype of the ALICE IROC has been built, commissioned and tested at the CERN Proton Synchrotron.
\subsection{IROC prototype}
\label{subsec-IROCprototype}
\begin{figure}
\centering
\includegraphics[width=0.5\textwidth, clip]{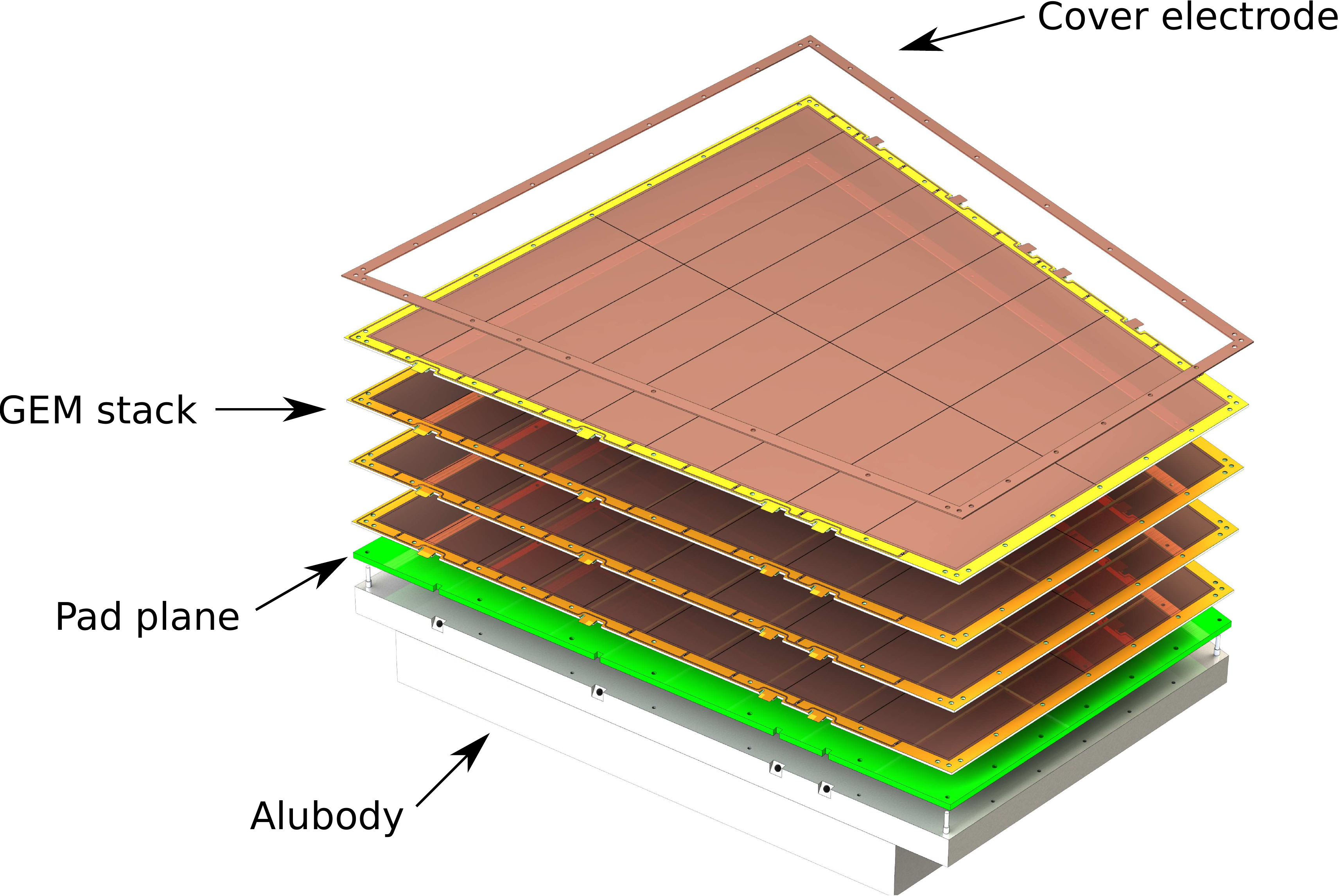}
\caption{Exploded view of the IROC prototype with the individual components being depicted.}
\label{fig-IROCexp}
\end{figure}

The 4 GEM IROC prototype is assembled on a spare IROC chamber body \cite{NIMA_TPC}. It is a trapezoidal chamber with a size of $497 \times (292 - 467)$ \si{\milli\meter\squared}. The mechanical structure of the chamber consists of four main components: an aluminum frame (alubody), a \SI{3}{\milli\meter} Stesalit insulation plate, the pad plane (5504 pads of $4 \times 7.5$ \si{\milli\meter\squared}) and the GEM stack with the cover electrode on top. Figure \ref{fig-IROCexp} shows an exploded view of the chamber.
A set of four large-size GEM foils in the baseline configuration has been produced in the CERN MPGD workshop with single-mask technique and prepared according to the procedures as described in \cite{TDR_TPCU}.
The cover electrode is a \SI{1}{\milli \meter} thick PCB that surrounds the active area of the GEM foils and assures the homogeneity of the drift field at the borders of the chamber.

The chamber is mounted in a test box, which contains a drift cathode and a rectangular field cage with outer dimensions of $57 \times 61$ \si{\centi\meter\squared}. This allows for the application of the nominal drift field of \SI{400}{\volt/\centi\meter} over a drift length of \SI{11.5}{\centi\meter}.

The detector is operated with a Ne-CO$_2$-N$_2$ (90-10-5) gas mixture. 

\subsection{Experimental setup}
\label{subsec-setupPS}
The \dEdx{} performance of the IROC prototype is evaluated in a test beam at the Proton Synchrotron (PS) at CERN. The prototype is installed in the PS East Area experimental hall (T10 beam line). The T10 beam line allows for the adjustment of the momentum of the secondary beam ranging from 1 to 7\,GeV/$c$ \cite{PS_doc}. For the purposes of this performance study, a beam containing electrons and negative pions with a momentum of 1\,GeV/$c$ is chosen.
For triggering and beam definition, a set of two scintillators is used. A threshold Cherenkov Counter provides a reference measurement for particle identification.

\subsection{Readout}
\label{subsec-readoutPS}
The IROC GEM prototype is equipped with 10 EUDET front-end cards (FEC), which were borrowed from the LCTPC (Linear Collider TPC) collaboration. This allows for the readout of about 1200 channels, covering a \num{6}-\SI{7}{\centi\meter} wide corridor over the whole length of the chamber. The FECs are based on the PCA16 / ALTRO chips \cite{PCA16, ALTRO}. The ALTRO chip was originally designed for the ALICE experiment. The RMS noise of the readout system has been determined to be about 600 electrons. A zero suppression threshold of 2 ADC counts was used, corresponding to around 2000 electrons (\SI{120}{\nano\second} peaking time and \SI{12}{\milli\volt/\femto\coulomb} conversion gain). The sampling frequency was set to \SI{20}{\mega\hertz}. 

The similarity of the FECs to the ALICE FECs allows for the usage of the current TPC readout system \cite{NIMA_TPC}: The data is read out via the backplane with two Readout Control Units (RCU) and transferred via optical links to a Local Data Concentrator PC, which contains the ReadOut Receiver Card (RORC) and runs the ALICE data acquisition system DATE \cite{NIMA_TPC}. The data from the two scintillators, the Cherenkov counter and a sensor monitoring the ambient conditions are read out via the CAMAC system. The two data streams are synchronized based on an event tag.

The average DAQ rate was $\sim$ 300 events/spill with a spill length of \SI{0.5}{\second}, whereas the beam rate was of the order of 2000 particles/spill.

\subsection{d\textit{E}/d\textit{x} measurements}
\label{subsec-dEdx}
The study of the particle identification capabilities of the detector prototype is conducted with AliRoot \cite{Aliroot}, the framework for reconstruction, simulation and analysis in ALICE.

Only events with single tracks are selected in order to make use of the particle identification provided by the Cherenkov counter. Additional cuts are applied on the number of clusters per track and the cluster drift time. Moreover, tracks with clusters at the acceptance edge are rejected.
\begin{figure}
\centering
\includegraphics[width=0.5\textwidth, clip]{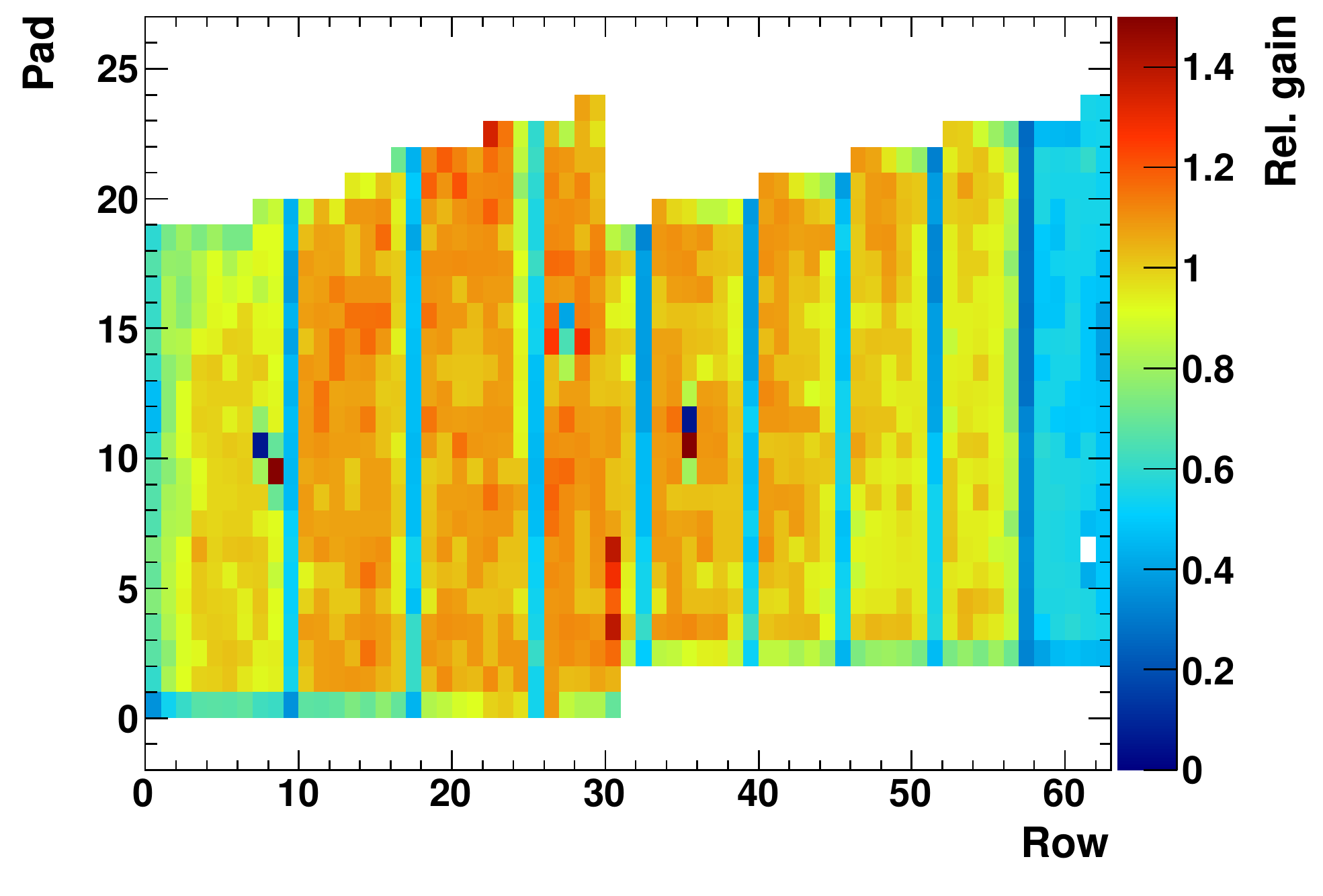}
\caption{Normalized gain map. For further details see text.}
\label{fig-gainmap}
\end{figure}

Geometric imperfections of the GEM system as well as channel-by-channel variations of the front-end electronics may cause variations of the gain. In order to minimize the impact of such variations, the normalized gain is extracted for each pad individually with tracks from the test beam. Figure \ref{fig-gainmap} shows the resulting gain map. Clearly visible are pad rows with lower gain, which are due to an overlap with supporting structures of the GEM stack. Also visible is a low-gain region for pad rows >57 due to a malfunctioning HV segment in one of the foils. For rows not affected by the mentioned effects, the observed gain spread is $\sim$ \SI{10.5}{\percent}. In order not to bias the \dEdx{} measurement, the charge of individual clusters is corrected by the normalized gain factor on a pad-by-pad basis. 

The specific energy loss \dEdx{} is then given by the truncated mean of the \SI{70}{\percent} lowest cluster charges $Q_{tot}$ of the ionization clusters of a track, which are reconstructed on the 63 pad rows on the readout anode. The approach of the truncated mean is chosen in order to symmetrize the energy loss distribution, that can then be fitted with a Gaussian function to obtain its mean value $\langle \mathrm{d}E/\mathrm{d}x \rangle$ and width $\sigma(\mathrm{d}E/\mathrm{d}x)$. The relative \dEdx{} resolution is then defined as $\sigma(\mathrm{d}E/\mathrm{d}x)/\langle \mathrm{d}E/\mathrm{d}x \rangle$.
\begin{figure}
\centering
\includegraphics[width=0.5\textwidth, clip]{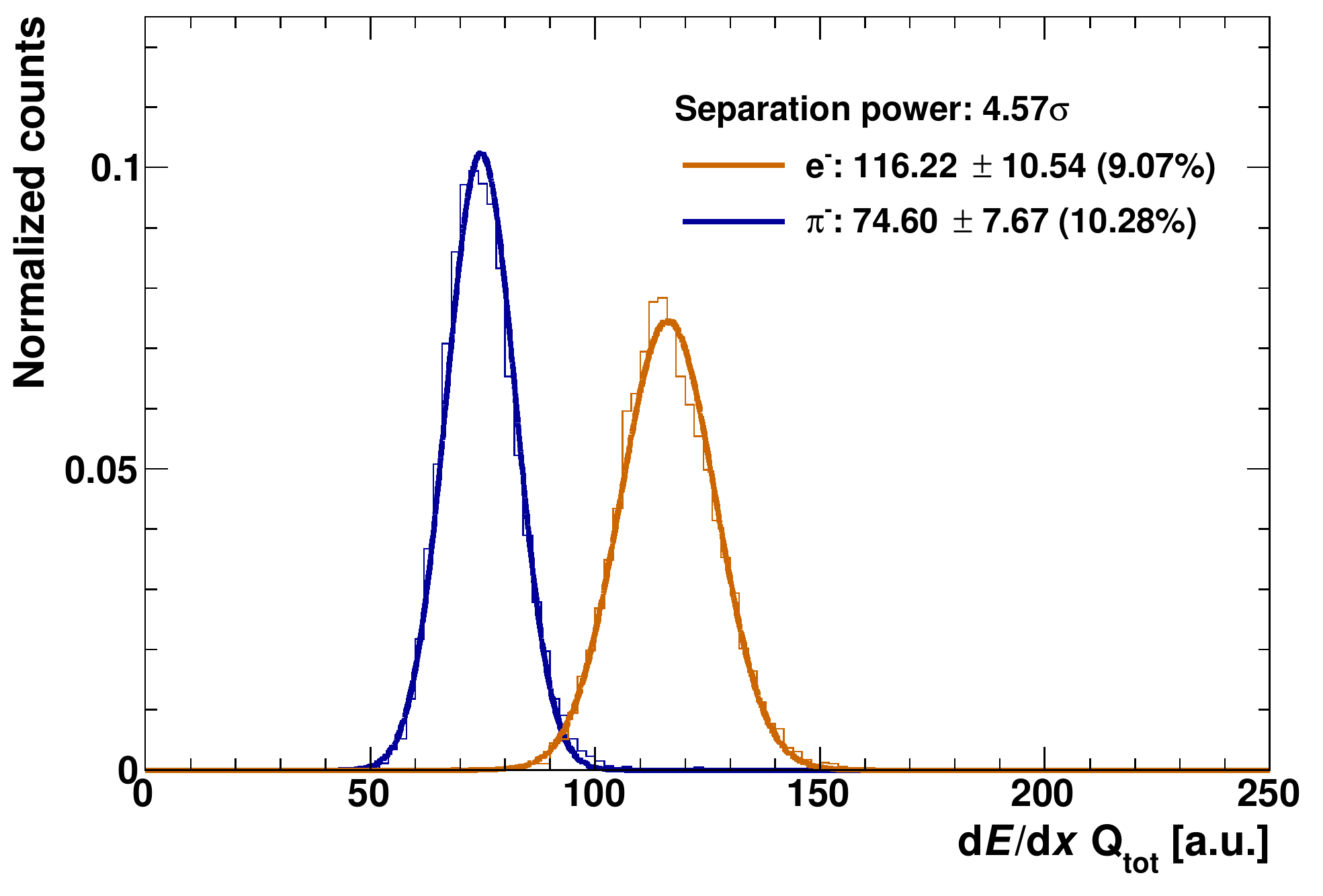}
\caption{Truncated (0–70 \%) energy loss distributions for electrons (\textit{orange}) and pions (\textit{blue}) with 1\,GeV/$c$ momentum measured with the IROC GEM prototype at a gain of 2000.}
\label{fig-dEdxQtot}
\end{figure}
Figure \ref{fig-dEdxQtot} shows an exemplary \dEdx{} spectrum of 1\,GeV/$c$ electrons and pions recorded at a gain of about 2000 and depicts the corresponding separation power, defined as

\begin{equation}
S_{AB} = \frac{2 \lvert \langle \mathrm{d}E/\mathrm{d}x \rangle_A - \langle \mathrm{d}E/\mathrm{d}x \rangle_B \rvert }{\sigma(\mathrm{d}E/\mathrm{d}x)_A + \sigma(\mathrm{d}E/\mathrm{d}x)_B}
\end{equation}
for two particle species $A$ and $B$. 

\begin{figure}
\centering
\includegraphics[width=0.5\textwidth, clip]{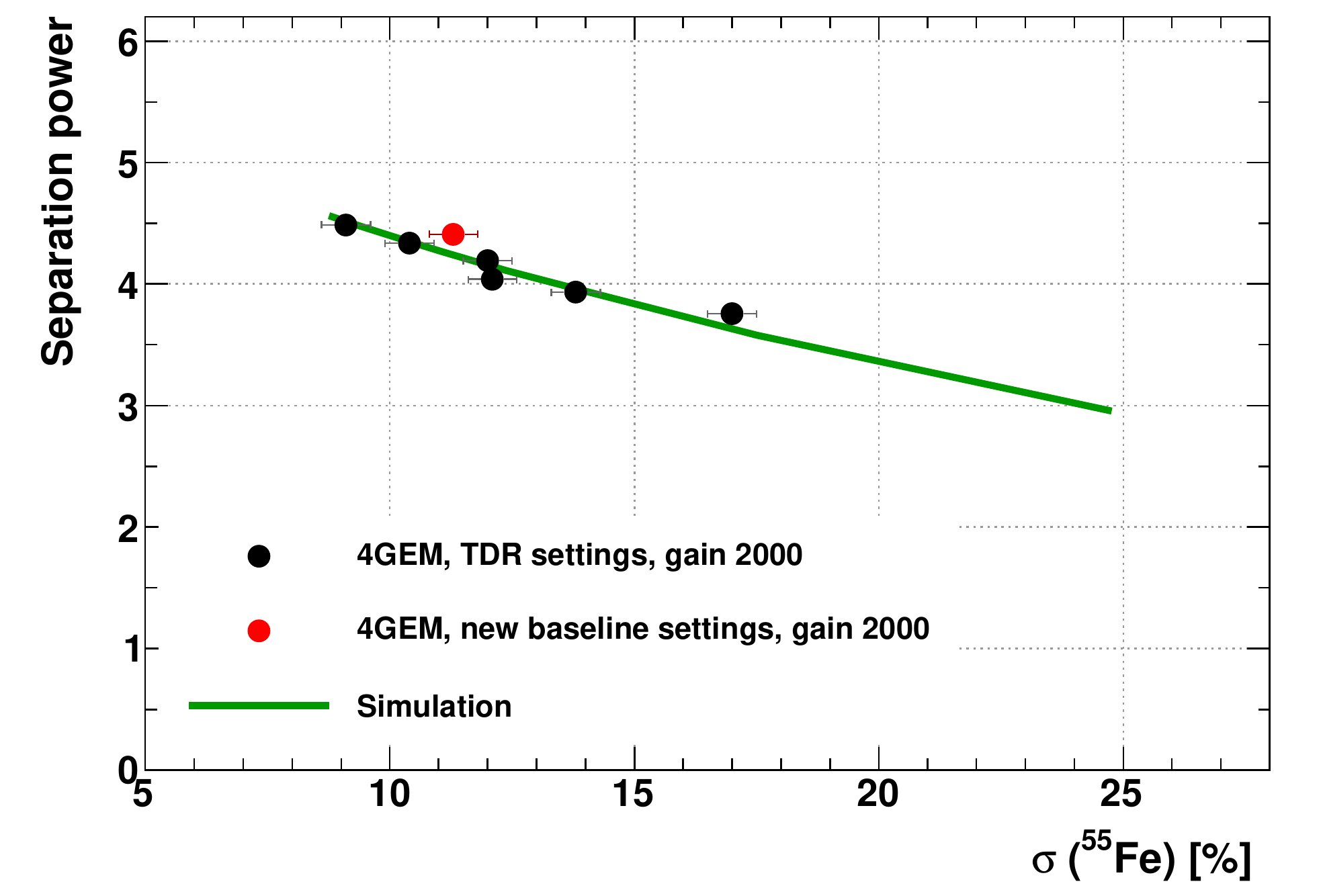}
\caption{Separation power of pions and electrons with a momentum of 1\,GeV/$c$ as a function of the $^{55}$Fe resolution at a gain of 2000. The green curve shows the result of a simulation. Reproduced from \cite{TDR_TPCU-add}.}
\label{fig-SigmaSep}
\end{figure}

A set of different voltage settings was tested during the test beam. The respective voltage settings, taken from measurements carried out with small prototypes \cite{TDR_TPCU}, are characterized in terms of the resolution of the $^{55}$Fe peak and ion backflow at a gain of 2000. 

The performance of the IROC prototype is shown as a function of the $^{55}$Fe resolution in Fig. \ref{fig-SigmaSep}. As expected, the separation power decreases as the $^{55}$Fe resolution is degraded. The measurement is compared to a microscopic simulation, in which the effective primary electron efficiency of the readout system is varied. This leads to a degradation of both the $^{55}$Fe and the \dEdx{} resolution, which reproduces the data very well.

The \dEdx{} resolution measured with the GEM IROC under a voltage configuration optimized for the specific requirements of the ALICE TPC is fully compatible with that of the MWPC IROC.

Concluding, the current \dEdx{} resolution of the ALICE TPC and thus its excellent PID capabilities will be preserved after the upgrade.

\section{Summary and outlook}
\label{sec-outlook}
The ALICE TPC will be upgraded for the LHC RUN 3 and beyond to operate at a \SI{50}{\kilo\hertz} rate in \PbPb{} collisions, demanding for a non-gated and continuous readout. Employing a stack of four GEMs with different hole pitches, the requirements of the readout system in terms of ion backflow, energy resolution and stability against discharges are met.

The \dEdx{} performance of a large-sized prototype equipped with such a readout structure has been evaluated at the CERN PS and demonstrated, that the PID capabilities of the TPC will not be compromised after the upgrade.


\begin{thebibliography}{99}
\bibitem{ALICE} K.~Aamodt {\it et al.} [ALICE Collaboration], "The ALICE experiment at the CERN LHC", JINST {\bf 3} (2008) S08002. http://dx.doi.org/10.1088/1748-0221/3/08/S08002

\bibitem{ALICE_UPGR} B.~Abelev {\it et al.} [ALICE Collaboration], "Upgrade of the ALICE Experiment: Letter of Intent", CERN-LHCC-2012-012 (2012). https://cds.cern.ch/record/1475243

\bibitem{NIMA_TPC} J.~Alme {\it et al.}, "The ALICE TPC, a large 3-dimensional tracking device with fast readout for ultra-high multiplicity events", Nucl. Instrum. Meth. A \textbf{622} (2010) 316. http://dx.doi.org/10.1016/j.nima.2010.04.042

\bibitem{GEM} F. Sauli, "GEM: A new concept for electron amplification in gas detectors"  Nucl. Instrum. Meth. A \textbf{386} (1997) 531. http://dx.doi.org/10.1016/S0168-9002(96)01172-2

\bibitem{TDR_TPCU-add} J.~Adam {\it et al.} [ALICE Collaboration], "Addendum to the Technical Design Report for the Upgrade of the ALICE Time Projection Chamber", CERN-LHCC-2015-002 (2015). https://cds.cern.ch/record/1984329

\bibitem{TDR_TPCU} B.~Abelev {\it et al.} [ALICE Collaboration], "Upgrade of the ALICE Time Projection Chamber", CERN-LHCC-2013-020 (2013). https://cds.cern.ch/record/1622286

\bibitem{PS_doc} D.~J.~Simon {\it et al.}, "Secondary beams for tests in the PS East experimental area", PS-PA-EP-Note-88-26 (1988). http://cds.cern.ch/record/1665434

\bibitem{PCA16} A.~Kaukher, O.~Schäfer, H.~Schröder, and R.~Wurth, "Status of TPC-electronics with Time-to-Digit Converters", EUDET-Memo-2009-08 (2009). http://www.eudet.org/e26/e28/e42441/e78196/EUDET-MEMO-2009-08.pdf

\bibitem{ALTRO} R.~Esteve Bosch, A.~Jimenez de Parga, B.~Mota and L.~Musa, "The ALTRO Chip: A 16-Channel A/D Converter and Digital Processor for Gas Detectors", IEEE Transactions on Nuclear Science \textbf{50} (2003) 2460. http://dx.doi.org/10.1109/TNS.2003.820629

\bibitem{Aliroot} AliRoot - ALICE Offline, http://aliweb.cern.ch/Offline/AliRoot/Manual.html.

\end{thebibliography}
\end{document}